\documentstyle[preprint,aps,prl,epsfig]{revtex}

\begin{document}
\newcommand{\sinf}{\raisebox{-.7ex}{$\stackrel{<}{\sim}$}}
\newcommand{\ssup}{\raisebox{-.7ex}{$\stackrel{>}{\sim}$}}

\ifx\undefined\psfig\def\psfig#1{    }\else\fi
\ifpreprintsty\else
\twocolumn[\hsize\textwidth
\columnwidth\hsize\csname@twocolumnfalse\endcsname       \fi    \draft
\preprint{  }
\title{Spin-based optical quantum gates via Pauli blocking in
semiconductor quantum dots}
\author{E. Pazy,$^{1}$ E. Biolatti,$^{1,2,3}$ T. Calarco,
$^{4,5}$ I. D'Amico,$^{1,2}$ P. Zanardi,$^{1,2}$ F.
Rossi,$^{1,2,3}$ and P. Zoller$^4$}
\address{$^1$ Institute for Scientific Interchange
(ISI), I-10133 Torino, Italy \\$^2$ Istituto Nazionale per la
Fisica della Materia (INFM), Torino, Italy\\ $^3$ Physics
Department, Polytechnical University of Torino,
I-10129 Torino, Italy\\
$^4$ Institute for Theoretical Physics, University of Innsbruck,
A-6020 Innsbruck, Austria\\
$^5$ INFM (Trento) and ECT*, I-38050 Villazzano (TN), Italy}
\maketitle

\begin{abstract}  We present a solid-state implementation of
ultrafast conditional quantum gates.
Our proposal for a quantum-computing device is based on the
 spin degrees of
freedom of electrons confined in semiconductor
 quantum dots, thus benefiting from relatively long decoherence times.
More specifically, combining Pauli blocking effects with properly tailored
ultrafast laser pulses, we are able to obtain sub-picosecond spin-dependent
switching of the Coulomb interaction, which is the essence of our conditional
 phase-gate proposal. This allows us to realize
{\it a fast two qubit gate which does not translate into fast
decoherence times} and paves the road for an all-optical spin-based quantum computer.
\end{abstract}
\ifpreprintsty\else\vskip1pc]\fi \narrowtext

Recent advances in quantum-information
science~\cite{NielsenChuang} have led to a number of schemes for
implementing quantum information processing (QIP) devices. Several
theoretical proposals have been made and admirable experimental
progress has been made in quantum optics~\cite{Wineland,Haroche},
NMR~\cite{Cory}, and solid state proposals including Josephson
junctions and quantum
dots~\cite{Schoen,Loss98,Imamoglu99,Biolatti,Lukin,Bayer,Magnets}.
Quantum dot (QD) implementation schemes based on the electronic
spin degrees of freedom~\cite{Loss98,Imamoglu99} are characterized
by relatively long decoherence times, compared to charge
excitations~\cite{Biolatti}, which on the other hand provide much
stronger interactions and thus faster gate operation.

Below we propose a novel implementation scheme for the realization
of a fast two-qubit gate which does not translate into fast
decoherence time. While our qubit is the spin of an excess
electron in a QD, the two-qubit gate relies on interaction between
excitons. Thus we merge ideas from both the fields of spintronics
and optoelectronics: using spin as quantum memory, and charge for
the interaction between qubits, we can benefit (i) from the
``low'' spin decoherence rates of conduction electrons in III-V
semiconductors~\cite{Kikkawa97}, and (ii) from ultrafast
(sub-picosecond) optical gating of charge excitations. Combination
with innovative schemes for all optical coherent control of the
electron spin~\cite{Imamoglu99,Gupta01} sets the stage for an all
optical spin-based QIP.

We consider an excess electron in a semiconductor QD, its spin
degrees of freedom defining the qubit $|0\rangle$, $|1\rangle$.
Our quantum register consists of an array of GaAs-based QDs each
dot containing one excess conduction electron~\cite{Loss98}.
Single-qubit operations can be done
optically~\cite{Imamoglu99,Gupta01} or on a pico-second time scale
by employing g-factor modulated materials or time-dependent
magnetic fields~\cite{Loss98}. The key ingredient in any quantum
computation scheme is the two qubit gate, that together with
single qubit operations forms a universal set of
gates~\cite{Lloyd96}.

We aim at implementing a two-qubit phase gate, defined through the
truth table $|\alpha \rangle \otimes|\beta\rangle \rightarrow
e^{\imath \alpha\beta \theta} |\alpha \rangle
\otimes|\beta\rangle$ where $\alpha,\beta\in\{0,1\}$. The dynamics
required to perform the gate operation exploits a {\em
Pauli-blocking} mechanism, as observed experimentally in
QDs~\cite{Bonadeo98,Chen00}. We assume that the QDs can be
individually addressed via laser excitation, using, e.g.,
near-field techniques~\cite{Imamoglu99} or energy-selective
addressing of QDs~\cite{Biolatti}. The control of the phase
accumulated by Coulomb interactions is obtained in the following
way: by shining a $\sigma^{+}$ polarized laser pulse on the QD,
due to the Pauli exclusion principle a $|M_J^e=-1/2 ,M_J^h=+3/2
\rangle$ electron-heavy hole pair is created in the s-shell only
if the excess electron
--- already present in the QD --- has a spin projection $1/2$. Thus
with a $\pi$-pulse we obtain an exciton conditional to the spin
state (qubit)
\begin{equation} \label{swap}
\alpha|0\rangle+\beta|1\rangle\to\alpha|0\rangle+\beta|x^-\rangle,
\end{equation}
where the logical states $|0\rangle$ and $|1\rangle$ shall be
defined below, and the charged exciton state $|x^-\rangle$ plays
the role of an auxiliary state needed to obtain the desired phase.
In this way one obtains precise spin control of the switching on
(and off) of further Coulomb interactions. In particular, this
allows the switching on of exciton-exciton interactions on
neighboring quantum dots conditional to the spins (qubits) being
in state $|1\rangle \otimes |1\rangle$. We emphasize that the
presence of the photo-generated electron-hole pair is only
required during the gating, after which the latter will be
annihilated via a second laser pulse. Thus excitonic interactions
can be switched on and off, in contrast to proposals where qubits
are stored in excitonic states~\cite{Biolatti}.

For a quantitative description, we consider the Hamiltonian of two
interacting QDs  $H =\sum_\nu H_\nu+H_{ab}+\sum_\nu H^{\rm
int}_\nu$, where the index $\nu$ runs over the QD label
$a,b$~\cite{Jacak98}. The single-QD Hamiltonian is $H_\nu=
H_\nu^{c}+ H_\nu^{cc}$, where the term describing the
noninteracting dynamics of the electrons and holes confined within
the QDs is $H_\nu^{c} = \sum_{i,\sigma=\pm 1/2}
\epsilon_{i,\sigma}^e c_{\nu,i,\sigma}^{\dagger} c_{\nu,i,\sigma}
+ \sum_{j,\sigma'=\pm 3/2} \epsilon_{j,\sigma'}^h
d_{\nu,j,\sigma'}^{\dagger} d_{\nu,j,\sigma'} $, and subscripts
$i$ and $j$ denote single-particle states --- e.g.,
$c_{\nu,i,\sigma}^{\dagger}$ ($d_{\nu,i,\sigma}^{\dagger}$) is a
creation operator for a conduction band electron (hole) in the
$i$-th single particle state of QD $\nu$, with spin projection
$\sigma$. The Coulomb-interaction part of the Hamiltonian
$H_\nu^{cc}$ is then composed of three terms: electron-electron,
electron-hole, and hole-hole interactions.  The coupling of the
carrier system with a classical light field is $H_\nu^{\rm int}= -
\sum_{ij} \left [ \mu_{ij}^{eh} E^{*}(t)e^{-\imath
\omega_{L}t}c_{\nu,i,-{1/2}}^{\dagger}d_{\nu,j,{3/2}}^{\dagger} +
h.c. \right]$, where $E(t)$ is a $\sigma^{+}$ circularly polarized
high frequency laser field, with a central frequency $\omega_{L}$,
and $\mu_{ij}^{eh}$ denotes the dipole matrix element between the
electron and hole wave functions.

An idealized model for two neighboring QDs is obtained by
considering dynamics restricted to the Hilbert space of relevant
states, which are the two spin states of the ground-state excess
electron $|0\rangle_\nu$, $|1\rangle_\nu$ representing the qubit,
and the auxiliary excitonic quantum states $|x^-\rangle_\nu$
($\nu=a,b$). These states are defined as eigenstates of the single
dot Hamiltonians $H_\nu$. In a single-particle picture these
states are approximated by $ |0 \rangle_\nu \equiv
c_{\nu,0,-1/2}^{\dagger} | \mbox{vac} \rangle$, $ |1 \rangle_\nu
\equiv c_{\nu,0,1/2}^{\dagger} | \mbox{vac} \rangle$, and $|x^{-}
\rangle_\nu \equiv c_{\nu,0,1/2}^{\dagger}
c_{\nu,0,-1/2}^{\dagger} d_{\nu,0,3/2}^{\dagger}| \mbox{vac}
\rangle$, while $| \mbox{vac} \rangle$ stands for the
electron-hole vacuum, i.e., the crystal ground
state~\cite{Biolatti,approxwavefunction}.

Our goal is now to design a process which starts from a
superposition of the qubit states, populates the auxiliary
excitonic states $|x^-\rangle$, and afterwards returns back to the
qubit space, achieving a phase gate. In the idealized model
outlined above, due to Pauli blocking, an exciton is obtained via
a $\sigma^+$-polarized laser pulse only if the excess electron is
in state $|1\rangle$ (see Fig. \ref{fig1}).
The presence of an exciton in both QDs induces the so-called
biexcitonic shift $\Delta E_{ab}$~\cite{Biolatti}, which is
defined as the difference between the
exciton binding energy in the presence or absence of a second
exciton in the neighboring QD. In this context, the most
straightforward gating strategy would be the following
\cite{Jaksch00}: (i) apply an exciting pulse to both QDs; (ii)
wait for a time $\Delta t \propto \theta/\Delta E_{ab}$, for the
phase $\theta$ to be accumulated, and (iii) bring the QDs to their
original state by a de-exciting pulse.
In this way the dipole-dipole interaction is active only if
the excess electrons in both QDs are in state $|1\rangle$ (see
Fig. \ref{fig2}), giving the desired state-dependence of the
two-qubit phase.
The typical time scale $\tau$ for a gate can be no shorter than
the inverse of the biexcitonic shift, which is of the order of 1 ps
\cite{Biolatti}.

The strongest decoherence source in charge-based gating schemes is
charge-phonon interaction. Recent  single dot photoluminescence
spectroscopy experiments~\cite{bayer} have measured ground state
excitonic phonon dephasing times of almost 1 ns, i.e. limited only
by the excitonic lifetime. Theoretical studies of pure dephasing
in small quantum dots~\cite{kuhn} have shown that this apparent
absence of dephasing is due to  polaronic effects which are manifested in 
the presence in the spectra of a
sharp zero-phonon line, and that the leading dephasing mechanism
is the coupling to acoustic phonons through the deformation
potential. The polaron formation time $t_0$ associated to this
mechanism is of the order of few picoseconds. In order to take
full advantage of the large strength of the zero phonon line
calculated in \cite{kuhn}, gating schemes based on charged states
in QDs, could then be performed adiabatically with respect to
$t_0$. In this way gating would be performed among dressed
long-lived quasi-particles (polarons) representing the
quasi-eigenstates of the total Hamiltonian. A technological
condition to be taken into account is that, since the needed laser
pulses are strongly energy selective (their typical energy
uncertainty is of $\sim 0.1$meV), a precise knowledge of the
characteristics of the QDs involved in the gate would be required.
An important point to be stressed is that, due to the strong
confinement regime, the typical energy associated to a phonon is
much smaller than the typical electron or hole intra-band level
spacing ($\sim 25$ meV), so that the coupling to phonons will not
induce a significant level renormalization and, in particular,
will not induce substantial mixing between different electronic
(hole) states. The relevant polaron states would then be well
approximated by the product between the ground electron (hole)
state and a linear superposition of all the relevant phonon states
\cite{kuhn}. This fact would allow us still to label each polaron
with the corresponding electron (hole) spin, i.e. it would still
be possible to apply, even to the dressed particles, a
Pauli-blocking based scheme.

Taking into account mixing between the heavy- and the light-hole
subbands, the eigenstate of the hole involved in our gate process
has to include a correction, weighted by the small parameter
$\varepsilon$, and becomes $\propto \varepsilon d_{0,1/2}^{\dagger
}\left| \mathrm{vac}\right\rangle +d_{0,3/2}^{\dagger }\left| \mathrm{vac}%
\right\rangle $. This will lead to a weak violation of the
selection rules for laser coupling, since a $\sigma
^{+}$-polarized laser in this case will not only excite the
desired transition $\left| 1\right\rangle \rightarrow \left|
x^{-}\right\rangle $ with Rabi frequency $\Omega $, but also the
unwanted one $\left| 0\right\rangle \rightarrow \left|
x^{+}\right\rangle \equiv
c_{0,-1/2}^{\dagger }c_{0,1/2}^{\dagger }d_{0,1/2}^{\dagger }\left| \mathrm{%
vac}\right\rangle $ with Rabi frequency $\varepsilon \Omega $.
Such an undesired exciton admixture for the qubit in state
$|0\rangle$ can be as big as 10\%. However, with an adiabatic
procedure involving a detuned laser pulse, we can guarantee that
this exciton population can be returned to zero after the laser
pulses, while the needed phase difference between the
computational basis states can still be obtained. Indeed, assuming
nonzero detuning, states $|0\rangle$ and $|1\rangle$ are
adiabatically connected to dressed states having different
excitonic components. Therefore each two-qubit computational basis
state $| \alpha \beta \rangle $ (where $\alpha ,\beta \in
\{0,1\}$) will still acquire, after the gate process, a different
phase $\phi _{\alpha \beta }$. Irrelevant single-qubit
contributions to the phase can be straightforwardly undone via
single-qubit rotations, recovering the transformation
$|\alpha\beta\rangle \rightarrow e^{\imath \alpha\beta \theta}
|\alpha\beta\rangle$, with the gate phase given by $\theta \equiv
\phi _{00}+\phi _{11}-\phi _{01}-\phi _{10}$. In order to
implement a universal set of quantum gates, $\theta\not=0$ has to
be obtained. To check whether this can be achieved in a realistic
situation, we need a model for the two-QD dynamics including hole
mixing. As anticipated above, the laser is assumed to be detuned
by a frequency $\Delta$ from the exciton transition. In a rotating
wave approximation the effective Hamiltonian after eliminating the
time dependence becomes
\begin{eqnarray}
\label{eq:effectiveh} H_{\mathrm{eff}}&=&\sum_{\nu =a,b}\left[
\frac{1}{2}\hbar\Omega (t)\left(| 1\rangle _{\nu }\langle x^{-}|
+\varepsilon | 0\rangle _{\nu }\langle x^{+}| \right)
+\mathrm{h.c.}\right]\nonumber\\
&&\mbox{}-\Delta\sum_{\nu =a,b;\sigma=+,-} | x^{\sigma }\rangle
_{\nu}\langle x^{\sigma }|\\
&&\mbox{}+\Delta E_{ab}\sum_{\sigma ,\sigma ^{\prime }=+,-}|
x^{\sigma }\rangle _{a}\langle x^{\sigma }| \otimes |x^{\sigma
^{\prime }}\rangle _{b}\langle x^{\sigma ^{\prime }}|,\nonumber
\end{eqnarray}
where $\Omega(t)=2\mu^{eh} E(t)/\hbar$ denotes the Rabi frequency.
As long as the Rabi frequency $\Omega$, the biexcitonic shift
$\Delta E_{ab}$ and the inverse pulse duration time scale
$\hbar\tau^{-1}$ are much smaller than the QD level spacing
$\Delta\epsilon\equiv\epsilon^e_1-\epsilon^e_0$, transitions to
excited states are negligible. Since our QDs are assumed to be in
the ``strong-confinement regime'', these conditions can be very
well satisfied. Typical experimental parameters employed in the
calculations below assume a width of the QDs (in the growth
direction) as well as the inter-dot barrier width of 50\AA\, and a
typical level spacing of $\Delta \epsilon \approx 25$\,meV. By
controlling the external in-plane electric field~\cite{Biolatti},
one is able to adjust the electrical dipole of the exciton which
effects the biexcitonic shift: for an electric field
$F=70$\,kV/cm, the biexcitonic shift is  $\Delta E_{ab}\approx
3$\,meV~\cite{Biolatti}.

The procedure we have in mind is as follows: We prepare our system
in a superposition of the logical states $\left| 0\right\rangle $
and $\left|
1\right\rangle $, and we switch on adiabatically a laser with detuning $%
\Delta $ from the excitonic transition. We want to follow
adiabatically the dressed energy levels of the system, up to a
point in which the ratio $\Delta /\Omega (t)$ becomes of the order
of unity or smaller. If we do so, the state $\left| 1\right\rangle
$ evolves into an adiabatic eigenstate with a bigger excitonic
component than the one that is reached starting from $\left|
0\right\rangle $. Then we switch the laser adiabatically off. If
we consider two quantum dots, this mechanism will let different
two-qubit computational basis states acquire a different phase
even in the presence of hole mixing. However, we must make sure
that at the end of gate operation no population is left in the
excitonic states. We can expect that this be the case as long as
adiabaticity is satisfied. To check this with a numerical
simulation, we chose a Gaussian Rabi frequency pulse shape
$\Omega(t)=\Omega_0e^{-(t/\tau)^2}$ in the Hamiltonian Eq.
(\ref{eq:effectiveh}). The adiabaticity condition takes then the
form $\Omega_0\tau\gg 1$. To this we must add the condition
$\tau\gg t_0$, as discussed above, in order to allow for the
formation of long-lived polaron states. The latter condition
turns out to be more stringent,
since the previous one can in principle be satisfied also for $\tau$ of
the order of picoseconds, by just increasing the Rabi frequency

Assuming a biexcitonic shift $\Delta
E_{ab}=3$ meV, a detuning $\Delta=\Delta E_{ab}$, a pulse width
$\tau=10$ ps and a peak Rabi frequency $\hbar\Omega_0\approx 5.9$
meV, we can obtain the desired value $\theta=\pi$ while the
population eventually left in the unwanted (single- and bi-)
exciton states remains of the order of $10^{-7}$. Results of the
simulation are shown in Fig. \ref{fig3}. This scheme has the
advantage of using the same pulse to excite both QDs, thus
relieving us from the need to tailor different pulses to address
the two QDs.

The fidelity of the described gate will be limited by deviations
from the idealized model (\ref{eq:effectiveh}), and decoherence
due to coupling to the environment, leading to energy relaxation,
exciton dephasing, spin decoherence~\cite{Kikkawa98,Imamoglu01}.
First, population leakage to other, e.g., excited states by the
laser is negligible for the present parameters since, as discussed
above, $\tau\Delta\epsilon/\hbar\gg 1$ . Hopping of the
exciton~\cite{Quiroga} between the QDs involved in the gate is not
allowed by spin selection rules, and should be negligible to other
QDs due to energy conservation arguments. Typical spin decoherence
times are of the order of $\sim 0.1\div 1 $ microseconds
\cite{Loss98,Imamoglu99}, and exciton lifetimes in the ground
state are given by the radiative life time of the order of
nanoseconds, while our gate can be performed on a few-ps time
scale. The probability for field ionization of the extra electron
in the QD ground state is proportional to the ratio between the
Stark shift $(eF)^2/m\omega_e^2$ and the height of the potential
well in the in-plane direction. Such a process should be
negligible for the present parameters. Dephasing due to coupling
of the electrons to the nuclear spin can be suppressed by applying
a further external magnetic field~\cite{Burkard99}. For the
simulations of Fig.~\ref{fig3} we have included decoherence
processes by employing a standard $T1-T2$ model, assuming a
dephasing time of 1 ns.

A quantum computer requires reading out the state of the qubit
(see, e.g., \cite{Imamoglu00}. This can be achieved easily by
adapting the familiar quantum jump technique from quantum optics
employing again the Pauli-blocking mechanism~\cite{Wineland}. We
emphasize that the same optical coupling to excitons provides a
mean to swap the spin-qubit to a photonic qubit by radiative
emission into a cavity, thus providing a natural optical
interconnect between spins as quantum memory and photonic qubits
for quantum communications~\cite{Cirac}.

In summary, we have proposed a two-qubit phase gate which benefits
from the vast time-scale separation between excitonic and spin
dephasing processes. Whereas our proposed qubit is given by the
spin of a conduction electron and thus decoheres on a micro-second
time scale, our conditional two-qubit phase gate is
driven/controlled by Coulomb interaction on a picosecond
time-scale. The ratio between gate operation time and the
coherence time of the quantum memory is therefore of the order of
$10^5$. Moreover, since our scheme employs the long-range characteristics
of Coulomb interactions, gating can be performed between
non-neighboring qubits. We demonstrated that our scheme tolerates a significant
amount of hole mixing (i.e., of violation of the Pauli-blocking
selection rules), under realistic parameter conditions. Further
improvement is expected from a more detailed engineering of the
decoherence sources, and this will be the subject of future
investigations.

Work supported by the Austrian Science Foundation, the European
Union (through the Programs IHP, IST-FET QIPC and ESF QIT), the
Institute for Quantum Information and NSF PHY99-07949.

\begin{figure}[b]
\begin{center}
\caption{\label{fig1}Quantum dots and energy
level scheme. Left: the excess electron is in state
$|-1/2\rangle\equiv|0\rangle$ and the transition induced by a
$\sigma^+$-polarized light is blocked. Right: the excess electron
is in $|+1/2\rangle\equiv|1\rangle$ and the exciton can be
excited.}
\end{center}
\end{figure}
\begin{figure}[htb]
\begin{center}
\caption{\label{fig2}Dynamics of the two-qubit
gate for the computational basis states
$|\alpha\rangle_a\otimes|\beta\rangle_b$
($\alpha,\beta\in\{0,1\}$). In the ideal case of perfect Pauli
blocking, the dipole-dipole interaction is present only for the
$|11\rangle$ component.}
\end{center}
\end{figure}
\begin{figure}[ht]
\begin{center}
\caption{\label{fig3}Upper panel: Biexcitonic population obtained
starting from an initial state $|00\rangle$ (short-dashed line),
$|01\rangle$ or $|10\rangle$ (solid line), and $|11\rangle$
(long-dashed line). Lower panel: Pulse shape and accumulated
phase.}
\end{center}
\end{figure}

\newpage
\begin{figure}[ht]
\begin{center}
\epsfig{figure=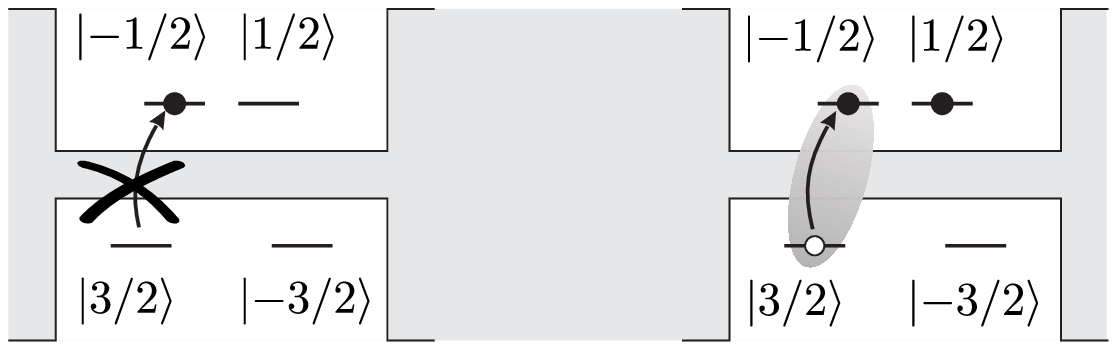,width=15truecm,angle=0}
\end{center}
\end{figure}

\newpage
\begin{figure}[htb]
\begin{center}
\epsfig{figure=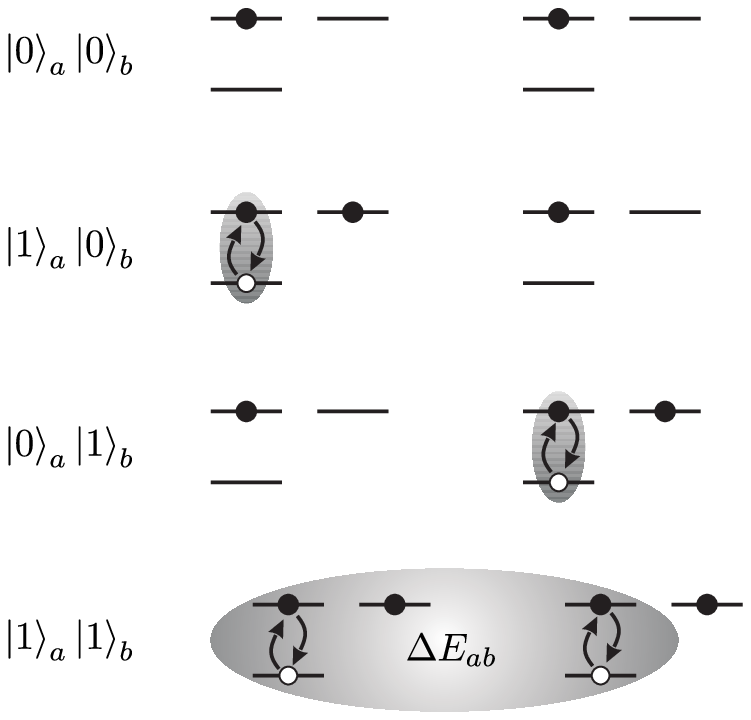,width=12truecm}
\end{center}
\end{figure}

\newpage
\begin{figure}[ht]
\begin{center}
\epsfig{figure=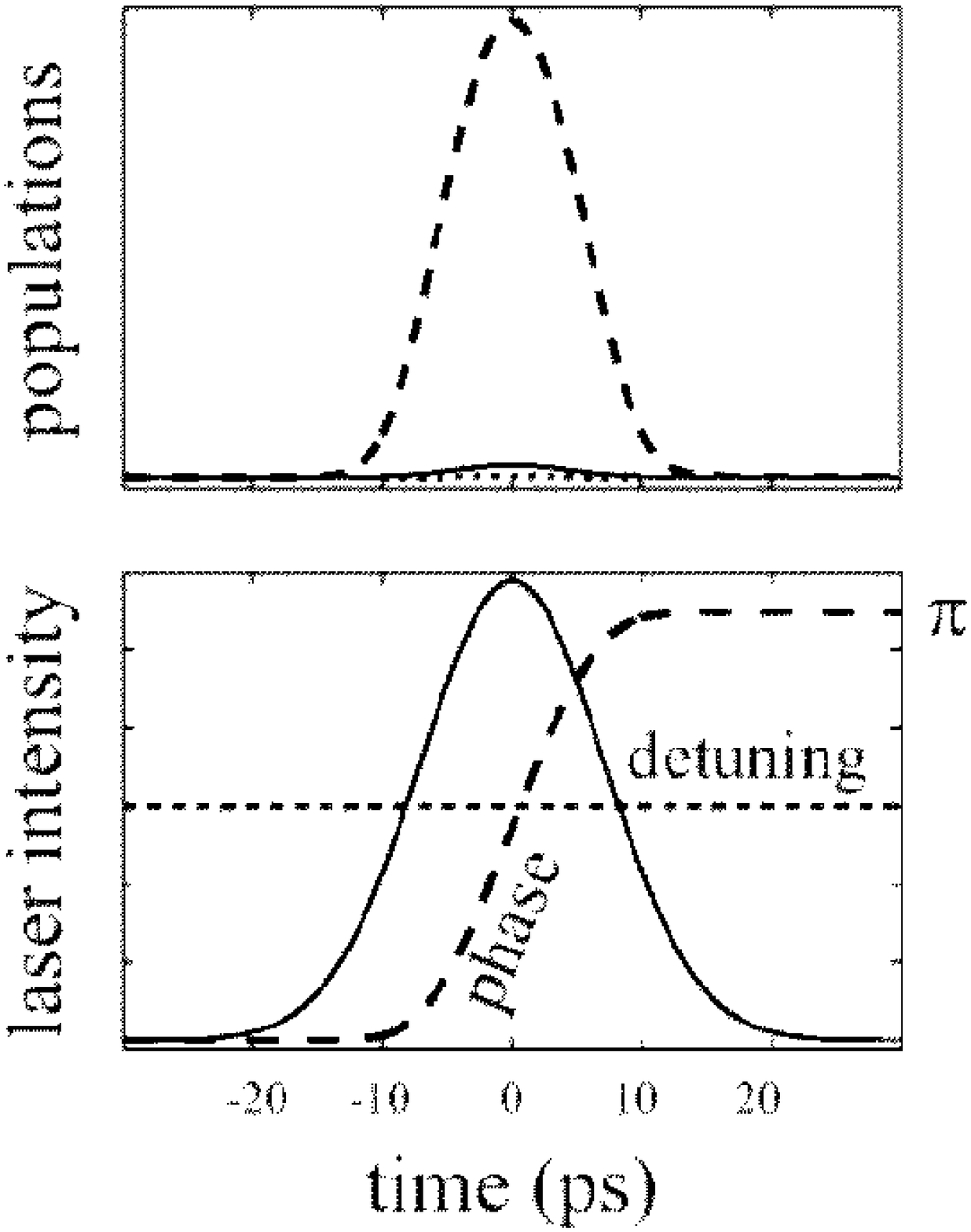,width=12truecm}\vspace{0.5truecm}
\end{center}
\end{figure}

\begin{thebibliography}{xxx}
\bibitem{approxwavefunction}
The charged exciton state was approximated as a product  of an
exciton and an electron single-particle wave function.
\end{thebibliography}
\end{document}